\documentclass[12pt]{article}
\usepackage{epsfig}

\def\beq{\begin{equation}}
\def\eeq#1{\label{#1}\end{equation}}
\def\eeqn{\end{equation}}

\def\beqa{\begin{eqnarray}}
\def\eeqa#1{\label{#1}\end{eqnarray}}
\def\eeqan{\end{eqnarray}}

\let\bar=\overbar

\def\Dslash{\not{\hbox{\kern-4pt $D$}}}
\def\dslash{\not{\hbox{\kern-2pt $\del$}}}

\def\msb{{\bar{\ssstyle M \kern -1pt S}}}

\usepackage{subfigure}

\def\Title#1{\begin{center} {\Large {\bf #1} } \end{center}}

\begin{document}

\Title{Semileptonic $B_s$ decays at Belle, Babar, and D0}

\bigskip\bigskip


\begin{raggedright}  

{\it Phillip Urquijo \index{Urquijo, P.}\\
University of Bonn\\
Bonn, GERMANY}
\bigskip\bigskip
\end{raggedright}

\noindent{\it Proceedings of CKM2012, the 7th International Workshop on the CKM Unitarity Triangle, University of Cincinnati, USA, 28 September - 2 October 2012}

\section{Introduction}

Semileptonic decays of $B$ mesons constitute a very important class of decays for determination of the elements of the Cabibbo-Kobayashi-Maskawa (CKM) matrix~\cite{ckm}, $|V_{ub}|$ and $|V_{cb}|$ and for understanding the origin of CP violation in the Standard Model (SM). An important aspect of semileptonic $B$ decays is that they only one hadronic current, and can therefore be described precisely by heavy quark theory, making them suitable for precision measurements of quark parameters.  The full semileptonic width is also large, making up more than 20\% of all $B$ decays.  The light, $B^0$, and $B^+$ meson decays have been precisely measured by experiment, with both exclusive and inclusive methods.  The experimental information on both the production and decay of  $B_s$ mesons, however is relatively limited.  The interest in the physics of the $B_s$ has intensified recently, particularly concerning studies of the dilepton asymmetry in $b \bar b$ production, at D0~\cite{D0_asl} and at LHCb \cite{LHCB_asl}.  They have also been used to precisely determine the $b-$quark production cross section~\cite{lhcb_crosssection}, and hadronisation fractions~\cite{lhcb_fs, tevatron_fs}.
Measurements of $B_s$ mesons may also provide precise input to heavy quark effective theory (HQET), and determination of the CKM matrix elements, as lattice predictions with heavy quarks ($s$ instead of $u$, $d$), and more phase space at lower recoil  lead to smaller theoretical uncertainties.

At the $B$-factories, data samples were collected with $e^+ e^-$ centre of mass (CM) energies of approximately $\sqrt{s}=10.87$ GeV, above the $B_s \bar B_s$ production thresholds. Babar and Belle collected approximately 3.2 fb$^{-1}$ and 121 fb$^{-1}$ near the $\Upsilon(5S)$, respectively.  The cross section of $\Upsilon(5S)$ production from $e^+e^-$ collisions at the $B$-factories, is approximately 30\% of $\Upsilon(4S)$ production. Furthermore only a fraction of $\Upsilon(5S)$ mesons decay to $B_s^0 \bar B_s^0$, $f_s = (19.9\pm3.0)\%$. The remainder proceed mainly to $B$ mesons.  This leads to approximately $B_s$ pair sample sizes of  $0.37\times10^6$ and $14\times10^6$ for Belle and Babar, respectively. 
Measurements of $B_s$ mesons, therefore face considerable statistical limitations with current $\Upsilon(5S)$ samples.
At the Tevatron, an abundant number of $B_s$ and other $b$-hadron species were produced.  The D0 experiment used a data sample of 1.3 fb$^{-1}$ of $p \bar p$ collisions to study exclusive semileptonic $B_s$ decays.

\section{Inclusive semileptonic $B_s$ decays}
An important prediction from HQET is that the decay widths of all semileptonic $B$ meson decays are equal, {\it i.e.} $\Gamma_{\rm sl } (B_u^+) = \Gamma_{\rm sl } (B_d^0) = \Gamma_{\rm sl } (B_s^0)$.   Non-perturbative QCD contributions to the decay width are modified, most significantly at third order in the spin orbit operator term \cite{bigietal, gronauetal}. Corrections from higher orders are expected to cancel these effects, leading to an overall correction of approximately 1\%. This implies that $SU(3)$ symmetry should hold.

The experimental signature for inclusive semileptonic $B_s$ decays is the presence of a single high momentum lepton. The other decay products of the $B_s$ are not exclusively reconstructed. A measurement of this quantity at $\Upsilon(5S)$ comprises of contributions from $B$ and $B_s$ decays, with a ratio of 4 to 1. It is therefore necessary to enhance the relative contribution from $B_s$ mesons, to minimise experimental systematic uncertainties.  To do this, tagging techniques have been employed that exploit the fact that $B_s$ decays proceed preferably via the Cabbibo favoured $B_s \to D_s^- X$ transition (${\cal B}(B_s \to D_s ^\pm X) = (93 \pm 25)\%$, and ${\cal B}(B \to D_s ^\pm X) = (8.3 \pm 0.8)\%$), followed by $D_s^- \to \phi X$. Although full $B_s$ reconstruction would be ideal for constraining background, as done at $\Upsilon(4S)$, the efficiencies are too low to be of use. Belle and Babar use slightly different methods based on these properties, described in turn below.

\subsection{Babar}
\begin{figure*}
\begin{center}
\subfigure[]{
\includegraphics[width=0.45\textwidth]{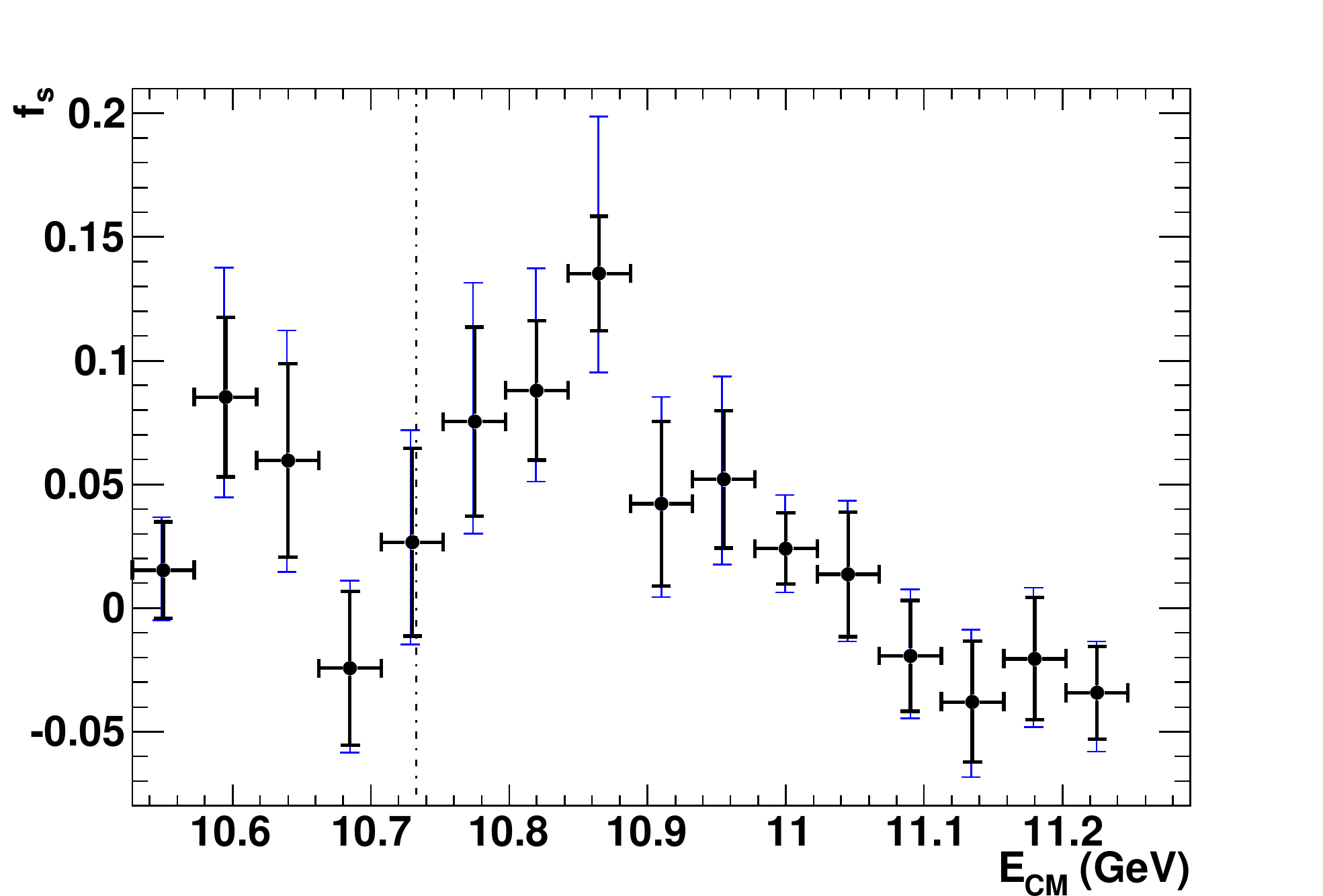}
\label{fig:fsfit}
}
\subfigure[]{
\includegraphics[width=0.45\textwidth]{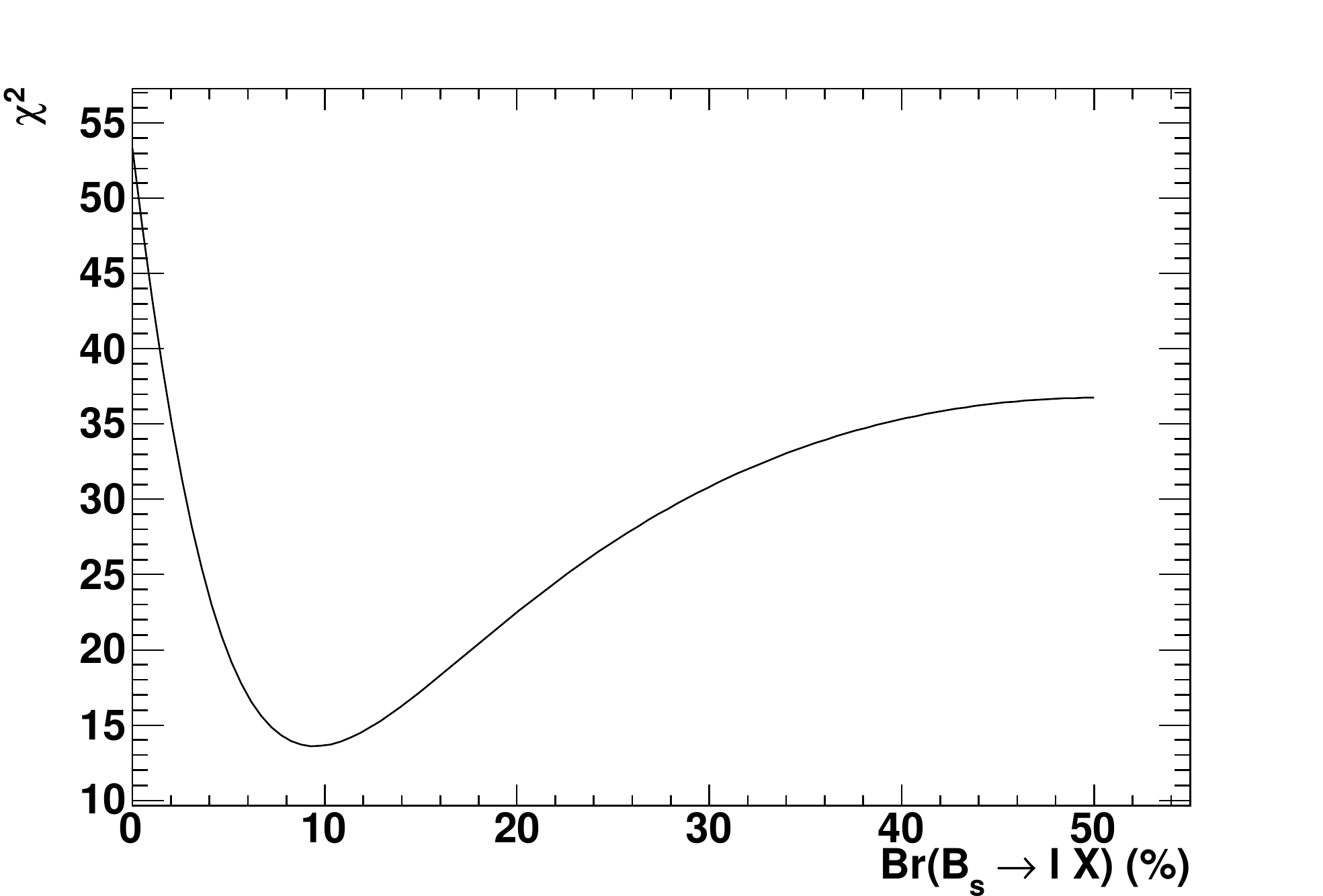}
\label{fig:chi2fit}
}
\end{center}
\caption{The Babar inclusive semileptonic decay measurement: (a) the fraction $f_s$ as a function of $E_{\rm CM}$, where the inner error bars show the statistical uncertainties, the outer error bars show the statistical and systematic uncertainties added in quadrature, and the dotted line denotes the $B_s$ threshold; (b) $\chi^2$ formed from the measured and expected yields, as a function of the semileptonic branching fraction; }
\end{figure*}

Babar searches for events that contain $\phi$ mesons (${\cal B}(D_s \to \phi X) = (15.7 \pm 1.0)\%$), which enhances the relative contribution of $B_s$ to $B$, by approximately 4.4. An additional high momentum lepton is searched for in the event, to identify semileptonic $B$ decays. This method clearly does not differentiate between $\phi$s produced from the same or the opposite $B_{(s)}$.  A third sample of events that contain at least one hadron in the final state is used to constrain some input parameters, most notably $f_s$. All three samples are measured as a function of the $e^+e^-$ CM energy.  The rates are defined as:
$C_h  = R_B [f_s \epsilon_h^s + (1-f_s)\epsilon_h]$, 
$C_\phi = R_B [f_s \epsilon_\phi^s P(B_s \bar B_s \to \phi X) + (1-f_s)\epsilon_\phi P(B \bar B \to \phi X)]$, and
$C_{\phi \ell} = R_B [f_s \epsilon_\phi^s P (B_s \bar B_s \to \phi \ell X) + (1-f_s) \epsilon_{\phi \ell} P(B \bar B \to \phi \ell X)]$, for the inclusive hadronic, $\phi$, and $\phi+\ell$ samples, respectively.
The $P(B\bar B \to \phi X, \phi \ell X )$ terms are the probabilities that a $\phi$ or $\phi+\ell$ pair is produced in a $B \bar B$ event, $\epsilon_i$ are reconstruction efficiencies, and $R_B$ is the production fraction normalised to Bhabha events.  Background from continuum is subtracted using data collected below the threshold for open $b$ hadron production. The measurement simultaneously extracts $f_s$ and ${\cal B}(B_s \to X \ell \nu)$. Values for $f_s$ are measured for each energy scan point (Fig. \ref{fig:fsfit}), and a global  $\chi^2$ is minimised over all scan points (Fig. \ref{fig:chi2fit}) to measure ${\cal B}(B_s \to X \ell \nu) = 9.5 ^{+2.5}_{-2.0}(stat.) ^{+1.1}_{-1.9}(sys.)$~\cite{babar_inclusive}. The dominant systematic uncertainty ($\sim$10\%), is due to the uncertainty on the inclusive $B_{(s)}\to D_s$ branching fractions. 

\subsection{Belle}
At Belle,  the inclusive semileptonic branching fraction (BF) is extracted from the ratio $N(D_s^+ \ell^+)/N(D_s^+)$, where $N(D_s^+)$ and $N(D_s^+ \ell^+)$ are the efficiency corrected quantities of $D_s^+$ mesons. $D_s$ mesons are reconstructed in the cleanest mode $D_s \to \phi \pi$, $\phi \to KK$, and fits to the invariant mass of the $KK\pi$ system are used to determine their yields.  Signal lepton candidates are required to be of the same electric charge as the reconstructed $D_s$ meson, to ensure that they both stem from different $B_s^0$ mesons. Background from continuum $c\bar c$ production, typically produces high momentum $D_s$ mesons, which are suppressed by the requirement $p^*(D_s^+)/p_{\rm max}^*(D_s^+)<0.5$, where $p^*$ denotes the momentum in the CM frame of the $e^+e^-$ beam. Residual background from $c \bar c$ continuum is subtracted using a $63$ fb$^{-1}$ sample collected at a CM of $\sqrt{s} = 10.52$ GeV, below the threshold of open $b$ hadron production. The background from secondary leptons (not coming directly from $B_{(s)}$ decays) and misidentified leptons, is estimated from a fit of the signal and background shapes derived from MC simulation, to the lepton momentum distribution.  
\begin{figure*}
\begin{center}
\subfigure[]{
\includegraphics[width=0.3\textwidth]{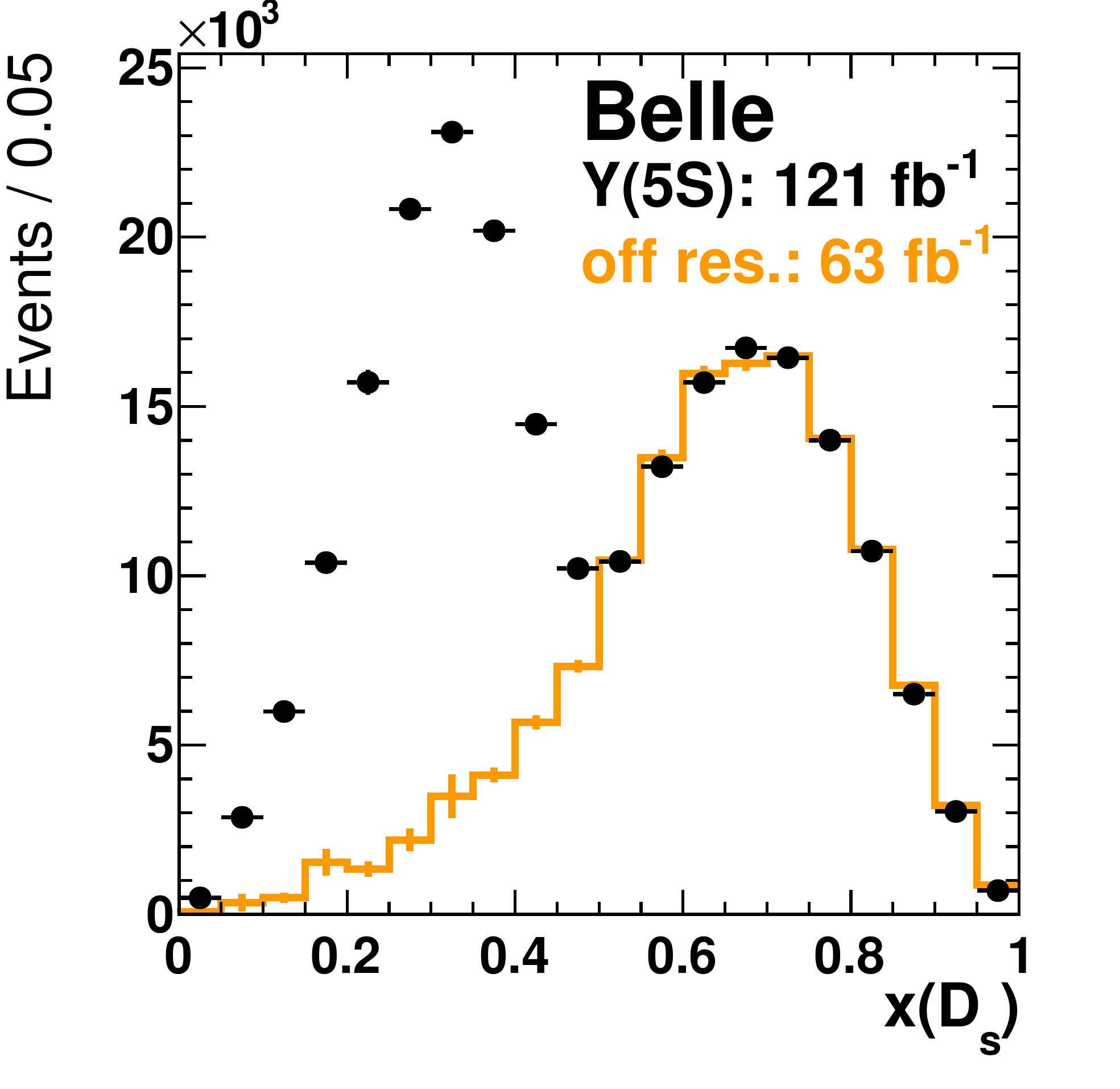}
\label{fig:xDs}
}
\subfigure[]{
\includegraphics[width=0.3\textwidth]{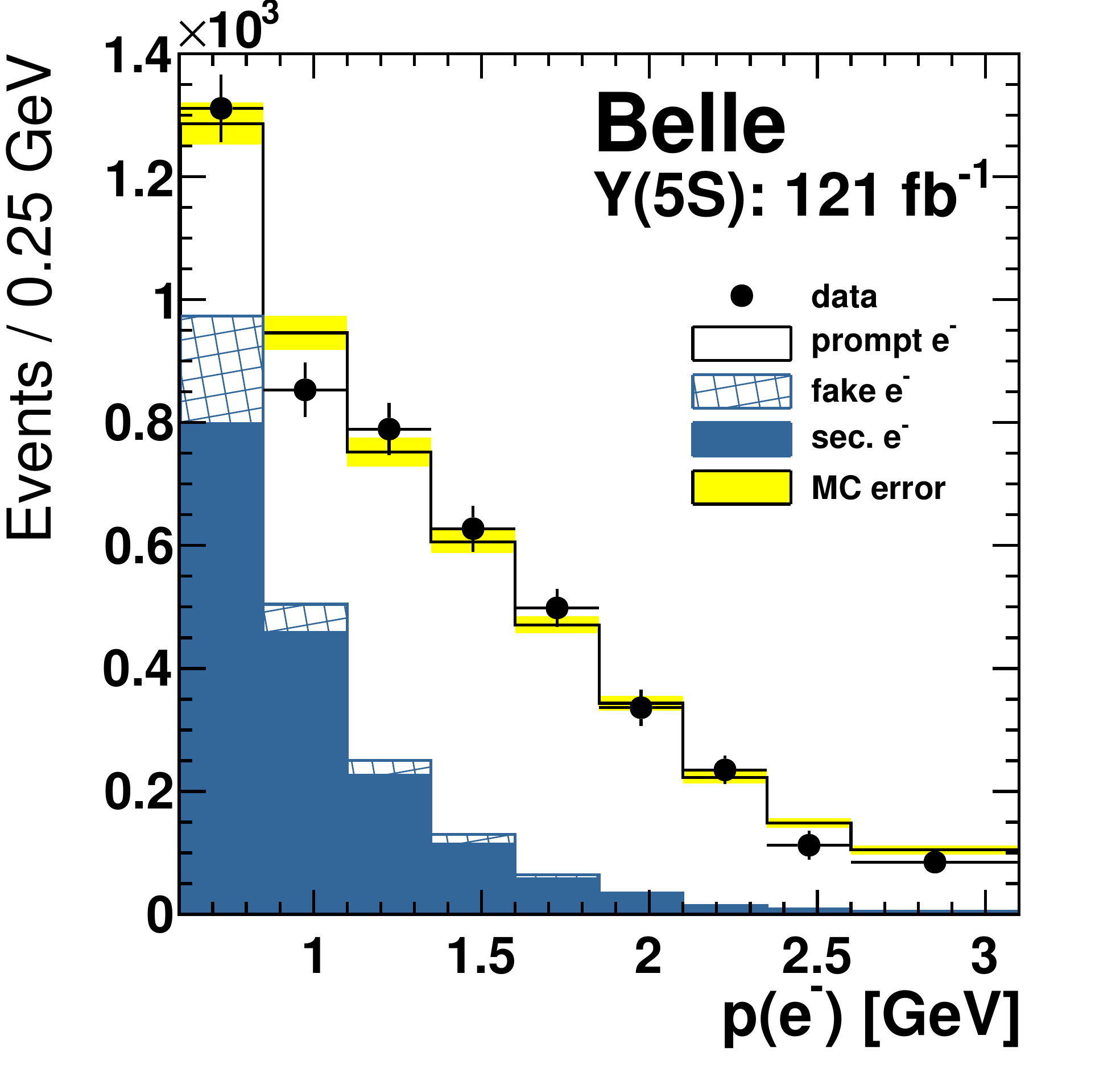}
\label{fig:elecmomfit}
}
\subfigure[]{
\includegraphics[width=0.3\textwidth]{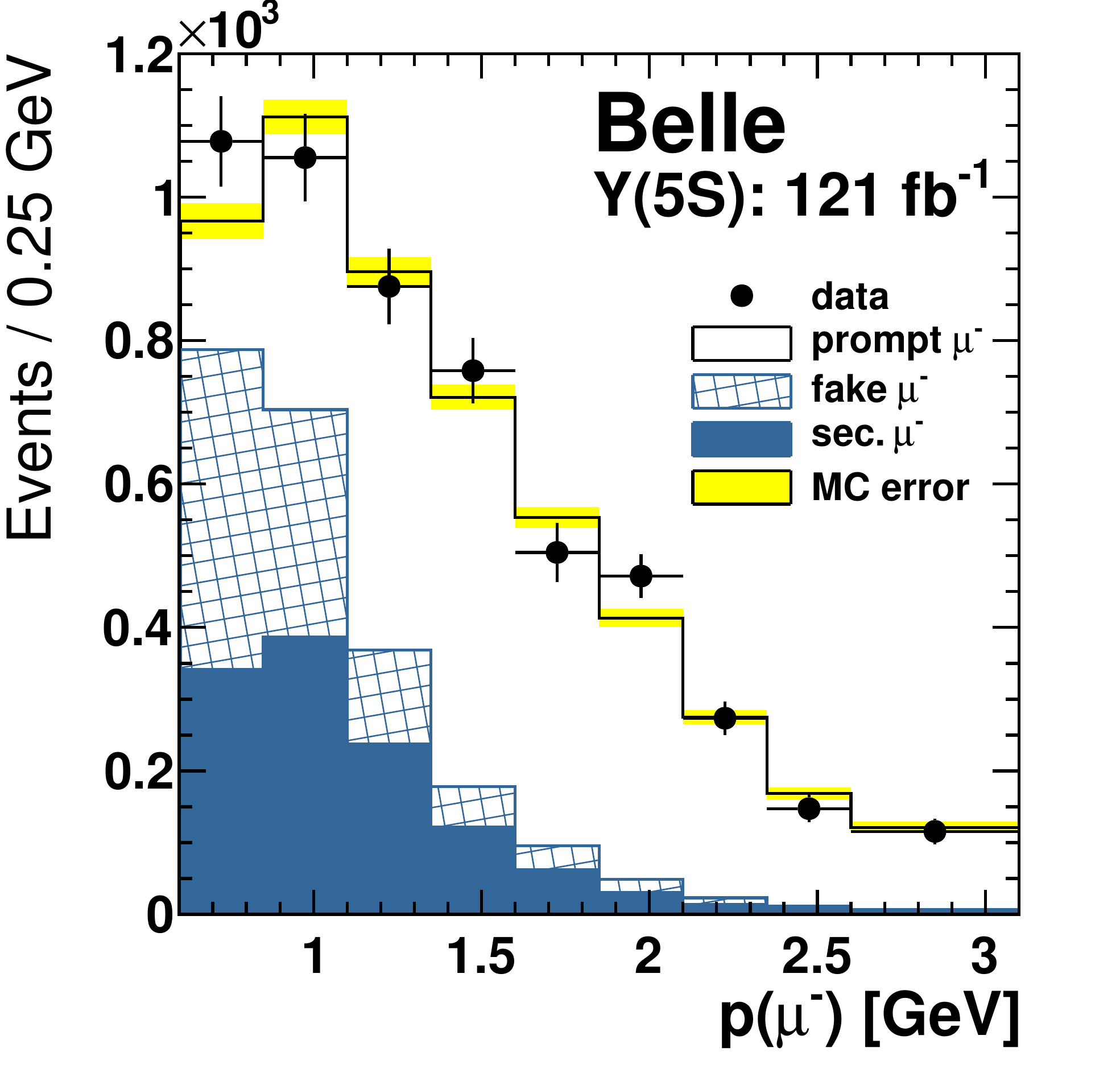}
\label{fig:muonmomfit}
}
\end{center}
\caption{The Belle inclusive semileptonic decay measurement. Shown are the momentum spectra obtained from $KK\pi$ mass fits: (a) In bins of $x(D_s^+)$
($D_s^+$ sample); (b)+(c) In bins of $p(e^+)$ and $p(\mu^+)$, respectively, where continuum
backgrounds have been subtracted using off-resonance data ($D_s^+\ell^+$ sample).
The MC uncertainty (yellow) comprises statistical and systematic uncertainties.}
\end{figure*}
The measured ratio  $N(D_s^+ \ell^+)/N(D_s^+)$ contains yields, ${\cal N}_s$ and ${\cal N}_{ud}$ , from $B_s$ and $B$ decays, respectively:
\begin{eqnarray}\label{nds}
\frac{N(D_s^+ \ell^+)}{N(D_s^+)}
=
\frac{{\cal N}_s(D_s^+ \ell^+)+{\cal N}_{ud}(D_s^+ \ell^+)}{{\cal N}_s(D_s^+)+{\cal N}_{ud}(D_s^+)},~~\ell=e,~\mu
\end{eqnarray}
The relative contributions from $B$ and $B_s$ to this ratio, explicitly defined on the right side of the equation, can be described in terms of known production fractions, known BFs (${\cal B}(B^{+/0}\to X\ell\nu)$ and ${\cal B}{(B_{(s)}\to D_{(s)}X)}$), and the signal of interest ${\cal B}(B_s \to X \ell \nu)$. The value for ${\cal B}(B_s \to X \ell \nu)$ can therefore be extracted from this relation.  The dominant systematic uncertainties introduced by the external parameters are $f_s/f_{u,d}$ (3.2\%), and the inclusive $D_s$ production BFs, $B_s\to D_sX$ (4.4\%) and $B \to D_s X$ (3.2\%).  
The final result is ${\cal B}(B_s \to X \ell \nu) = 10.6 \pm 0.5_{\rm stat.} \pm 0.7_{sys.}$ \cite{belle_inclusive}, where the errors are statistical, and systematic respectively.  The results are compatible with those from Babar, although much more precise, due to the 40 $\times$ larger data sample. This model independent measurement is the most precise absolute branching fraction of a $B_s$ decay.   Using the $B^0$ semileptonic width and the well measured $B_s$ and $B^0$ lifetimes, SU(3) symmetry is confirmed, {\it i.e.} $\Gamma_{\rm sl}(B_s^0) = (1.04 \pm 0.09)\cdot \Gamma_{
\rm sl}(B_d^0)$.

\section{Exclusive semileptonic $B_s$ decays}

The exclusive composition of the $B_s$ semileptonic decay width is not well known, however there have been some measurements of its parts.  The D0 and LHCb collaborations have measured BFs of two excited charm modes above the $D^0K$ threshold, leaving most of the decay width, {\it i.e.} $B_s \to D_s^{(*)} \ell^+ \nu_\ell$ still  to be explored.  $B_s$ meson semileptonic decays to heavier excited charm states provide us with important insight: they are sensitive to zero recoil dynamics and hence useful for testing heavy-quark effective theory.  $B_s$ decays may provide important insight into outstanding puzzles of HQET, most notably the $J=1/2$ versus $3/2$ puzzle. In the $B$ system, the narrow $J=1/2$ mesons appear to have significantly larger BFs than predicted by HQET. Experimentally it is very difficult to precisely isolate the wide $J=3/2$ components.   An alternative way of testing the HQET predictions for the width of $J=1/2$ versus $3/2$ decays is to perform the measurement in the $B_s$ system. All orbitally excited $D_s^{**}$ mesons have relatively narrow decay widths, including the $J=3/2$ mesons. Here we discuss the measurements by D0. More recent measurements by the LHCb collaboration are covered in a separate proceedings~\cite{bozzi}.

\subsection{D0}
\begin{figure*}
\begin{center}
\subfigure[]{
\includegraphics[width=0.4\textwidth]{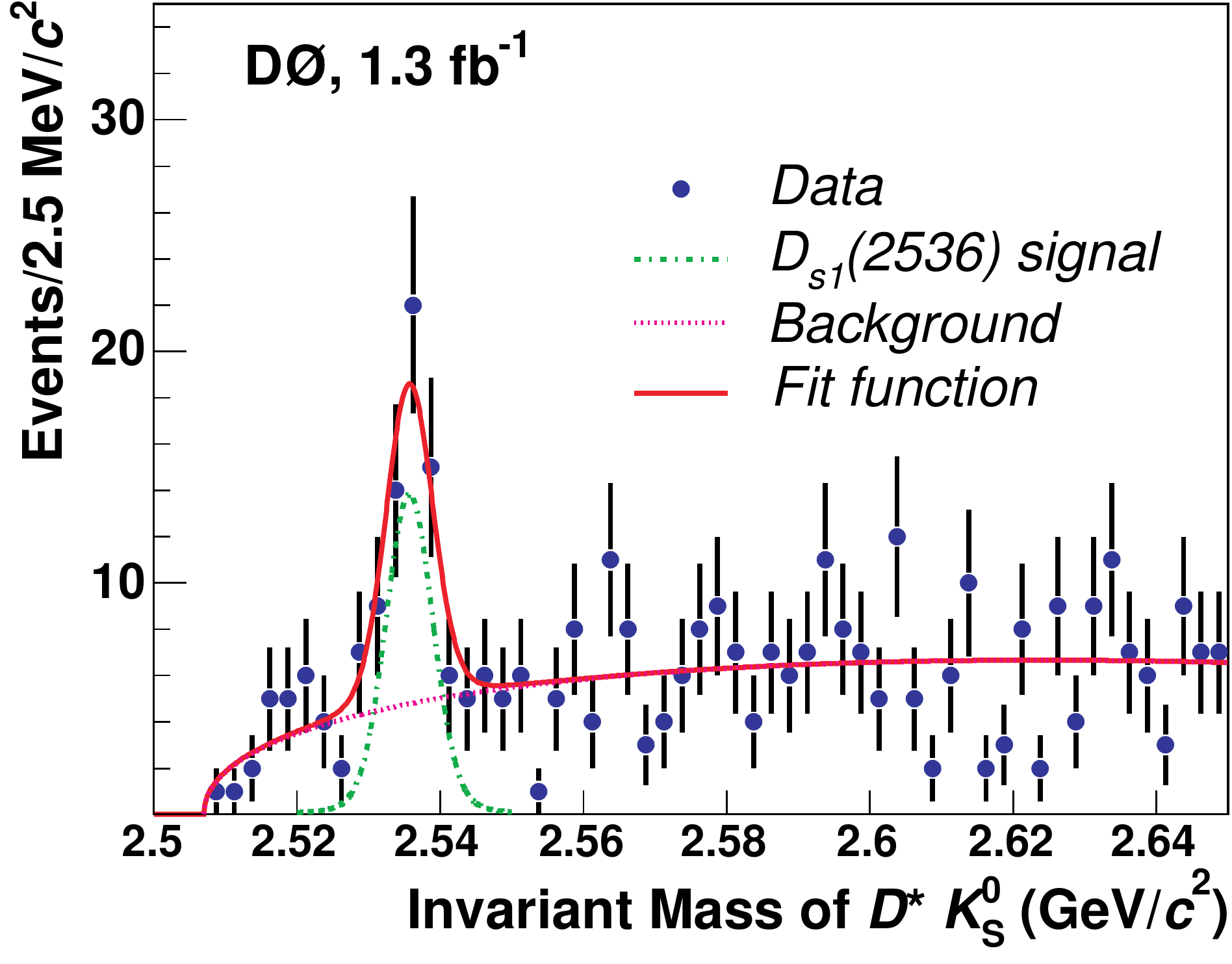}
\label{fig:D0a}
}
\subfigure[]{
\includegraphics[width=0.4\textwidth]{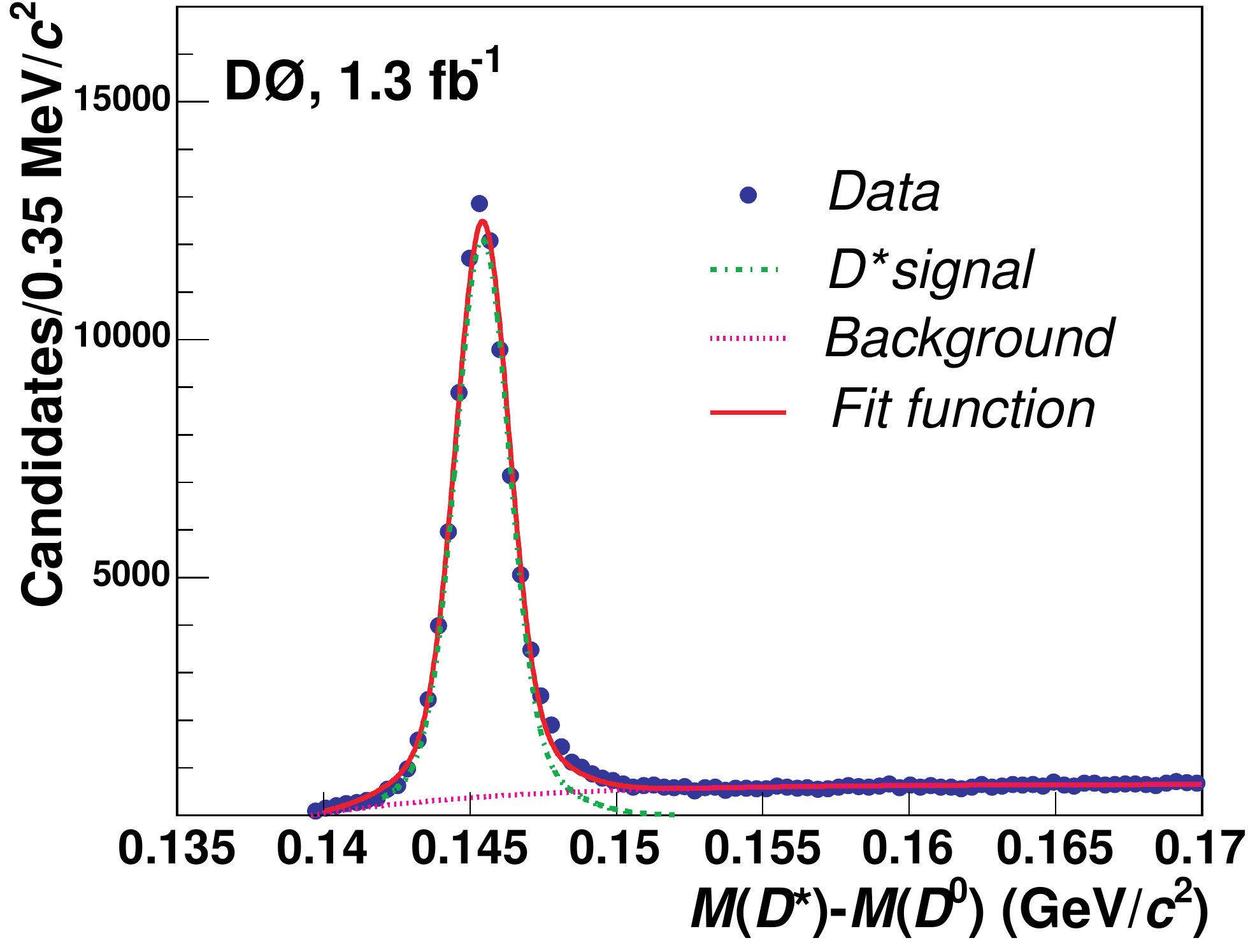}
\label{fig:D0b}
}
\end{center}
\caption{D0 exclusive semileptonic decay measurement: (a) fit to the invariant mass of $D^*K_S^0$ with an associated muon; (b) the mass difference $M(D^*)-M(D^0)$ for $b$ candidates with an associated muon.
}
\end{figure*}

The D0 experiment measured semileptonic $B_s$ decays into orbitally excited $P$-wave strange-charm mesons ($D_s^{**}$) \cite{tevatron_exclusive}.  In this decay, the $j_q=\frac{3}{2}$ angular momentum can be combined with the heavy quark spin to form the $J^P=1^+$, $D_{s1}$ state which must decay through a $D$ wave to conserve $j_q=\frac{3}{2}$. The $D_{s1}^\pm(2536)$ is expected to decay dominantly to a $D^*$ and $K$ meson to conserve angular momentum.  
The reconstructed decay chain was $b \to D_{s1}^-(2536)\mu^+\nu_\mu X$, $D_{s1}^-(2536) \to D^{*-}K_S^0$; $D^{*-}\to \bar D^0 \pi^-$, $K_S^0 \to \pi^+\pi^-$, $\bar D^0 \to K^+\pi^-$.  The branching fraction is normalised to the known value of ${\cal B}(\bar b \to D^{*-} \mu^+ \nu_\mu X)$, and the hadronisation fraction, $f_s$. 
A data sample corresponding to $1.3$ fb$^{-1}$ of integrated luminosity was used for the measurement. Figures \ref{fig:D0a} and \ref{fig:D0b} show the fit to the $D^* K_S^0$ invariant mass and the fit to the mass difference $m(D^*-D^0)$ in the normalisation mode, respectively. The resonance width was fixed to the known value from Babar.
The branching fraction product was determined to be 
$
f(\bar b \to B_s^0)\cdot {\cal B} (B_s^0 \to D_{s1}^- \mu^+ \nu_\mu X)\cdot {\cal B} (D_{s1}^- \to D^{*-} K_S^0) 
= [2.66 \pm 0.52 ({\rm stat.})\pm 0.45 ({\rm syst.})] \times 10^{-4}
$. The dominant errors arise from the $K_S^0$ reconstruction efficiency (11\%), the $b \to D^* \mu$ reconstruction efficiency (8.6\%), and ${\cal B}(b \to D^* \mu X$) (6.9\%).
Taking into account the measured value of $f_s$ and assuming ${\cal B}(D_{s1}(2536))=0.25$, the BF was found to be   $[1.03 \pm 0.20 ({\rm stat.})\pm 0.17 ({\rm syst.}) \pm 0.14 (f_s)] \%$, consistent with predictions from relativistic quark models with $1/m_Q$ corrections.

\section{Belle and Belle II prospects}
There is a good outlook for  semileptonic measurements near the $\Upsilon(5S)$ resonance with Belle and Belle II~\cite{belle2} data.  The estimated yields for Belle and Belle II are summarised in Table \ref{tab:belle2} for 3 signatures: ${\cal B}(B_s \to X \ell \nu)=11\%$, ${\cal B}(B_s \to D_s \ell \nu)=2.5\%$, and ${\cal B}(B_s \to K \ell \nu)=1.5\cdot 10^{-4}$. 
\begin{table}
\begin{center}
\begin{tabular}{l|cccccc}  
\hline
Tagging technique & Efficiency &  $N_{B_s}/N_B$ &  \multicolumn{3}{c}{Yields in 121 fb$^{-1}$(5 ab$^{-1}$)}\\ 
 &  &   & $X\ell \nu$ & $D_s \ell \nu$& $K \ell \nu$\\ 
\hline
 Untagged &   2     &     0.25      &     $2.7 \cdot 10^6$ & $7.2 \cdot 10^3$ & $2.5\cdot 10^3$  \\
 $\phi$ &   0.12     &     1             &     $1.6 \cdot 10^5$ & $450$ & $150$\\
 $D_s\to\phi\pi,K_sK,K^*K$ & 0.04 & 2.5 & $2.7 \cdot 10^4$$^\dagger$ & 140(6000) & 47 (2000) \\ 
 Semileptonic & 0.01 & $\gg 10$ & $6.8 \cdot 10^3$$^\dagger$ & 40(1500) & 12 (500) \\ 
 Full Reconstruction & 0.004 & $\gg 10$ & $5.4 \cdot 10^3$ & 15(620) & 5 (200) \\ 
 \hline
\end{tabular}
\caption{Belle II outlook. $\dagger$ denotes that opposite sign pairs must be ignored for the inclusive analysis. Full reconstruction and semileptonic tag efficiencies are based on measured $B^0$ equivalents.}
\label{tab:belle2}
\end{center}
\end{table}
With a sample of a few ab$^{-1}$, it would be possible to precisely measure the absolute BFs of most of the semileptonic decay modes at Belle II. However, measurements are not as straightforward as at the $\Upsilon(4S)$, due to contamination from $B^{+/0}$ decays, and to kinematic smearing from excited $B_s^*$ production. Although full $B_s$ reconstruction would mitigate $B^{+/0}$ background, such a technique will only be possible with Belle II. Leptonic and semileptonic tagging methods can be explored in the meantime.

\section{Conclusions}
Semileptonic decays are a standard candle for $B_s$ physics. They are being characterised using $\Upsilon(5S)$ data at the B-factories, complemented by studies of rarer exclusive modes at the hadron colliders.  The B-factories have tested SU(3) symmetry in  inclusive semileptonic $B_s$ decays, by precisely measuring the branching fraction to be $(10.5 \pm 0.5 {}^{+0.6}_{-0.7})$\% (private average). The excited charm mode $ B_s \to D_{s1} \mu\nu$ was studied by D0 and found to be in agreement with HQET with $1/m_Q$ corrections. The prospects for Belle and Belle II are promising to complete the survey of the semileptonic decay width, and to probe charmless semileptonic $B_s$ decays.


\begin{thebibliography}{99}

\bibitem{ckm}
N. Cabibbo, Phys. Rev. Lett. 10, 531 (1963); M. Kobayashi and T. Maskawa, Prog. Theor. Phys. 49, 652 (1973).


\bibitem{D0_asl} 
V. Abazov {\it et al.} [D0 Collaboration], Phys. Rev. D 82, 032001 (2010).

\bibitem{LHCB_asl} 
  Z.~Xing [LHCb Collaboration],
  arXiv:1212.1175 [hep-ex].

\bibitem{lhcb_crosssection}
 R.~Aaij {\it et al.}  [LHCb Collaboration],
  Phys.\ Lett.\ B {\bf 694}, 209 (2010). 

\bibitem{lhcb_fs}
 R.~Aaij {\it et al.}  [LHCb Collaboration],
  Phys.\ Rev.\ D {\bf 85}, 032008 (2012).

\bibitem{tevatron_fs}
  T.~Aaltonen {\it et al.}  [CDF Collaboration],
  Phys.\ Rev.\ D {\bf 77}, 072003 (2008).

\bibitem{bigietal}
  I.~I.~Bigi, T.~.Mannel and N.~Uraltsev,
  JHEP {\bf 1109}, 012 (2011).

\bibitem{gronauetal}
  M.~Gronau and J.~L.~Rosner,
  Phys.\ Rev.\ D {\bf 83}, 034025 (2011).
  
\bibitem{babar_inclusive}
  J.~P.~Lees {\it et al.}  [BaBar Collaboration],
  Phys.\ Rev.\ D {\bf 85}, 011101 (2012).

\bibitem{belle_inclusive} 
  C.~Oswald, P. Urquijo {\it et al.}  [Belle Collaboration],
  arXiv:1212.6400 [hep-ex].
  	
\bibitem{bozzi} 
  C.~Bozzi,
  arXiv:1303.4219 [hep-ex].
  
\bibitem{tevatron_exclusive}
  V.~M.~Abazov {\it et al.}  [D0 Collaboration],
  Phys.\ Rev.\ Lett.\  {\bf 102}, 051801 (2009).

\bibitem{belle2} 
  T.~Aushev, W.~Bartel, A.~Bondar, J.~Brodzicka, T.~E.~Browder, P.~Chang, Y.~Chao and K.~F.~Chen {\it et al.},
  arXiv:1002.5012 [hep-ex].


\end{thebibliography}
\end{document}